\documentclass[american]{ws-procs9x6}  

\usepackage[
]{packages}

\makeatletter
\g@addto@macro\bfseries{\boldmath}
\makeatother

\definecolor{darkred}{rgb}{0.5,0,0}
\definecolor{darkgreen}{rgb}{0,0.5,0}
\definecolor{darkblue}{rgb}{0,0,0.5}
\definecolor{darkyellow}{rgb}{0.5,0.5,0}
\definecolor{darkcyan}{rgb}{0,0.5,0.5}
\definecolor{darkmagenta}{rgb}{0.5,0,0.5}
\hypersetup{%
  colorlinks=true,
  linkcolor=darkblue,
  citecolor=darkmagenta,
  urlcolor=darkgreen,
  pagebackref=true,
  hyperindex=true,
  bookmarksopen
}

\crefname{figure}{Fig.}{Figs.}                             
\Crefname{figure}{Figure}{Figures}

\usepackage{macros}

\DeclareRobustCommand{\refCite}[1]{%
  \begingroup
  \romannumeral-`\x  
  \setcitestyle{numbers}%
  Ref.~[\cite{{#1}}]%
  \endgroup\xspace%
}
\DeclareRobustCommand{\refsCite}[1]{%
  \begingroup
  \romannumeral-`\x  
  \setcitestyle{numbers}%
  Refs.~[\cite{{#1}}]%
  \endgroup\xspace%
}

\begin{document}

\title{Light-Meson Spectroscopy \\ at Lepto- and Hadroproduction Experiments}

\author{B. Grube$^*$}

\address{%
  Institute for Hadronic Structure and Fundamental Symmetries, \\
  Technische Universität München, Garching, Germany \\
  $^*$E-mail: bgrube@tum.de
}

\begin{abstract}
  The excitation spectrum of light mesons, which are composed of up,
  down, and strange quarks, is studied since decades.  However, it
  still holds a number of puzzles and surprises that provide new
  insights into the nature of the strong interaction.  Recent
  high-quality data samples from several experiments allow us to not
  only study the properties of established mesons with unprecedented
  precision but to also search for new states.  These searches aim in
  particular at resolving the question of the existence of so-called
  exotic states, such as four-quark states or states with excited
  gluon fields.  Since light mesons have often large widths and are
  overlapping, the mapping of their spectrum is challenging and
  requires large quantities of data on different production and decay
  modes.  The data are analyzed using a framework of interfering
  quantum mechanical amplitudes known as partial-wave analysis (PWA).
  Most excited meson states decay into multi-particle final states,
  for which the PWA requires extensive modeling of the dynamics of the
  final-state hadrons.  I will give an overview on ongoing
  experimental studies of light mesons and discuss possible
  interpretations.  I will also touch on novel analysis techniques and
  the prospects for future progress.
\end{abstract}


\bodymatter

\section{Introduction}

The precision measurement of the spectrum of light mesons is the aim
of several experiments.  The focus of these experiments lies in
particular on confirming and finding new highly excited states, on
completing SU(3)$_\text{flavor}$ multiplets, and on searching for
exotic mesons, \ie states that cannot be composed of (just)
\qqbarPrime.  These data provide important input for theory and
phenomenology and eventually help to better understand the nature of
confinement.  The current analyses in the field are driven not only by
the high-quality data from experiments, but also by the development
and application of advanced analysis techniques and of more rigorous
theoretical models for partial-wave analyses (PWA).

\section{Spin-Exotic Light Mesons}

Spin-exotic mesons have \JPC quantum numbers\footnote{Here, $J$~is the
  meson spin and $P$~and $C$~are the eigenvalues of parity and charge
  conjugation, respectively.} of $0^{--}$, $(2n)^{+-}$, or
$(2n + 1)^{-+}$ with $n \in \mathbb{N}$ that are forbidden for
$\ket{\qqbar}$~states in the non-relativistic limit.  In the
light-meson sector, so far three spin-exotic candidate states---all
with $\JPC = 1^{-+}$---have been claimed by
experiments:\cite{Tanabashi:2018zz} \PpiOne[1400], \PpiOne[1600], and
\PpiOne[2100].  The latter one has been observed only by the BNL E852
experiment and needs confirmation.

\subsection{$3\pi$ Final State}

Some of the experimental claims are controversial, in particular the
observation of the \PpiOne[1600] in the $\Prho\pi$ decay.  The BNL
E852 experiment claimed a \PpiOne[1600] signal in a PWA of \num{2.5e5}
$3\pi$ events diffractively produced by an \SI{18}{\GeVc} $\pi^-$~beam
on a proton target.  The PWA was performed in the range from
\SIrange{0.1}{1.0}{\GeVcsq} of the reduced four-momentum transfer
squared~$t'$ using a PWA model containing
21~waves.\cite{Adams:1998ff,Chung:2002pu}  However, in a later
analysis of a more than ten times larger data sample of \num{2.6e6}
$3\pi$ events no evidence for the \PpiOne[1600] was
found.\cite{Dzierba:2005jg}
However, the non-observation claim was based on data in the narrower
range \SIvalRange{0.1}{t'}{0.5}{\GeVcsq}.

\begin{figure}[tbp]
  \vspace*{-4ex}%
  \centering
  \subfloat[][]{%
    \includegraphics[width=0.33\textwidth]{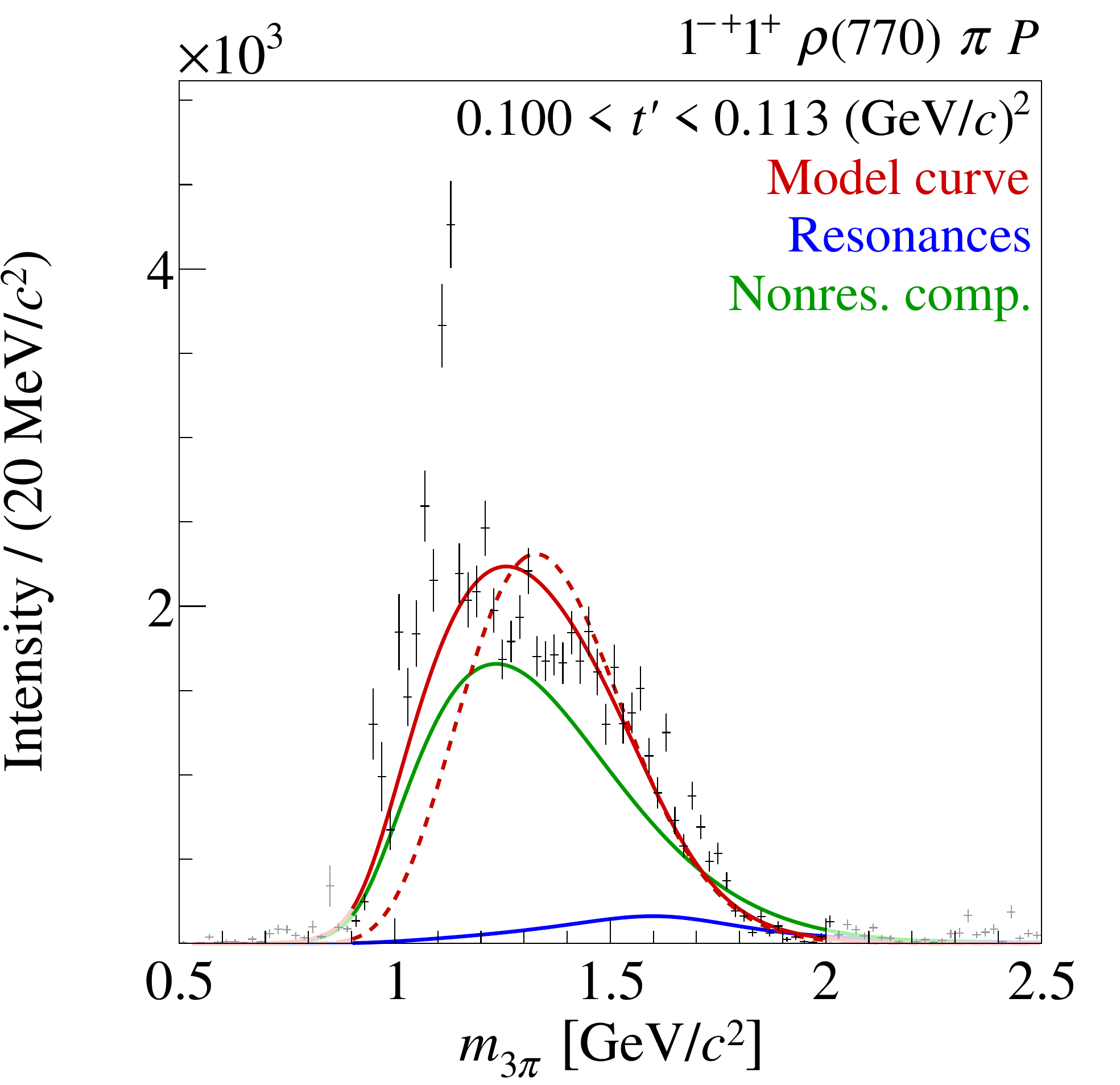}%
    \label{fig:intensity_1mp_tbin1}%
  }%
  \subfloat[][]{%
    \includegraphics[width=0.33\textwidth]{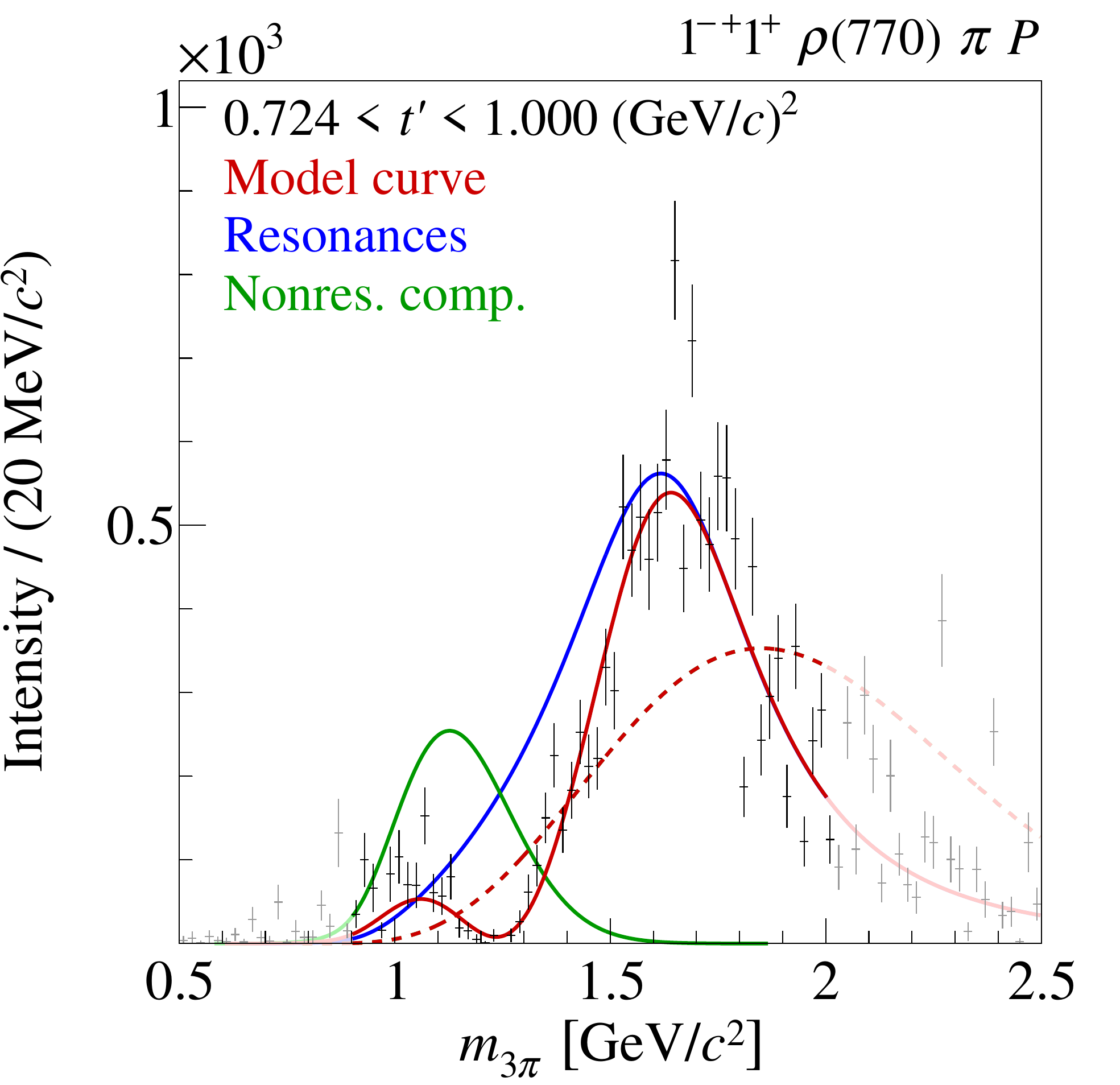}%
    \label{fig:intensity_1mp_tbin11}%
  }%
  \subfloat[][]{%
    \includegraphics[width=0.33\textwidth]{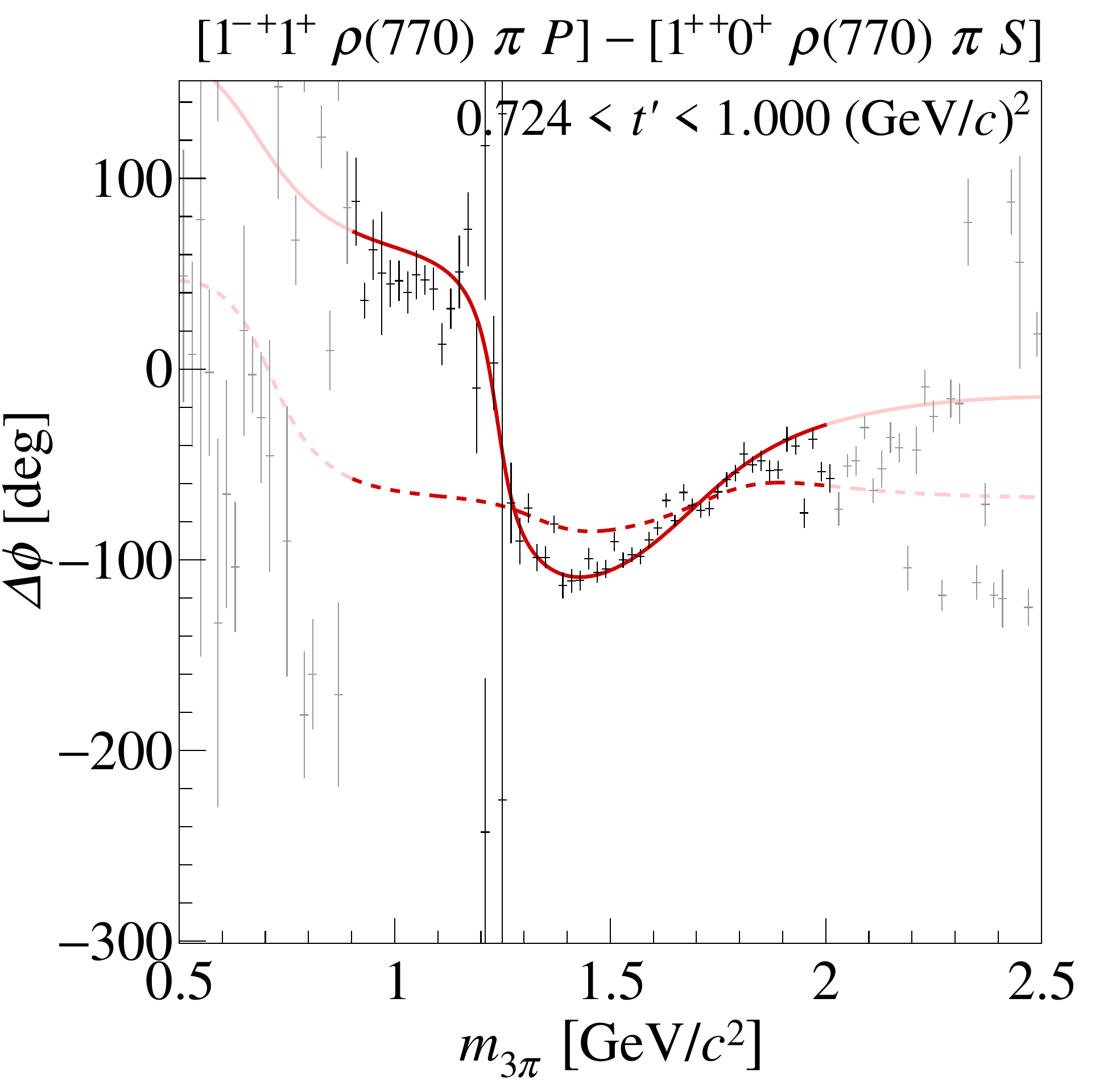}%
    \label{fig:phase_1mp_tbin11}%
  }%
  \caption{\subfloatLabel{fig:intensity_1mp_tbin1}~and~\subfloatLabel{fig:intensity_1mp_tbin11}:
    Intensity distributions of the $1^{-+}$ $\Prho\pi$ $P$-wave in the
    COMPASS \threePi proton-target data in the lowest and highest
    $t'$~bin, respectively.
    \subfloatLabel{fig:phase_1mp_tbin11}~Phase of this wave relative
    to the $1^{++}$ $\Prho\pi$ $S$-wave in the highest $t'$~bin. From
    \refCite{Akhunzyanov:2018lqa}.}
  \label{fig:1mp_3pi}
\end{figure}

Recently, the COMPASS experiment at CERN\cite{Abbon:2014aex}
published results of a PWA of a large data sample of \num{4.6e7}
$3\pi$ events diffractively produced by a \SI{190}{\GeVc} $\pi^-$~beam
on a proton target in the range
\SIvalRange{0.1}{t'}{1.0}{\GeVcsq}.\cite{Adolph:2015tqa,Akhunzyanov:2018lqa}
The employed PWA model contains 88~partial waves and is an extension
of the 36-wave set used in \refCite{Dzierba:2005jg}.  The PWA was
performed in 11~narrow $t'$~intervals.
\Cref{fig:intensity_1mp_tbin1,fig:intensity_1mp_tbin11} show the
measured intensity distributions (black points) of the spin-exotic
$1^{-+}$ $\Prho\pi$ $P$-wave for the lowest and the highest $t'$~bin.
Surprisingly, the shape of the intensity distribution changes
drastically with increasing~$t'$.  At low~$t'$, COMPASS observes a
broad distribution, whereas at high~$t'$, a peak emerges at
\SI{1.6}{\GeVcc}.  This suggests that the partial-wave amplitude has
large $t'$-dependent contributions from non-resonant processes.  The
continuous curves in \cref{fig:1mp_3pi} represent a resonance-model
fit, where the partial-wave amplitude is modeled as the coherent sum
(red curves) of a Breit-Wigner amplitude for the \PpiOne[1600] (blue
curves) and a non-resonant amplitude (green curves).  The modulation
of the intensity distribution with~$t'$ is well reproduced by the
model.  At low~$t'$, the intensity is described mostly by the
non-resonant component (green curve), whereas the peak at high~$t'$ is
described nearly completely by the Breit-Wigner component of the model
(blue curve).  The measured \PpiOne[1600] Breit-Wigner parameters are
$m_0 = \SIaerr{1600}{110}{60}{\MeVcc}$ and
$\Gamma_0 = \SIaerr{580}{100}{230}{\MeVcc}$.  That a resonance is
indeed required to describe the data, is demonstrated by a fit, where
the \PpiOne[1600] component is removed from the resonance model (red
dashed curves in \cref{fig:1mp_3pi}) leaving only the non-resonant
component in this wave.
Using the so-called freed-isobar PWA approach, it was also verified
that the observed \PpiOne[1600] signal is not an artifact caused by
inadequate isobar parameterizations (see
\refCite{Krinner:hadron2019} for details).

Performing the PWA and the resonance-model fit in narrow $t'$~bins
hence resolves the long-standing puzzle of the seemingly contradictory
results obtained by the BNL E852 experiment.  Using an even larger
wave set than the BNL analyses, the COMPASS results confirm that the
prominent peak reported in the first BNL analysis in
\refsCite{Adams:1998ff,Chung:2002pu} is an artifact caused by a too
small wave set.  The non-observation of the \PpiOne[1600] reported in
the second BNL analysis in \refCite{Dzierba:2005jg} is explained by
the exceptionally steep $t'$~dependence of the non-resonant component.
In the range $t' \lesssim \SI{0.5}{\GeVcsq}$, the non-resonant
component dominates and the COMPASS data show only a small
\PpiOne[1600] signal, which becomes prominent only for
$t' \gtrsim \SI{0.5}{\GeVcsq}$.  Hence the non-observation of the
\PpiOne[1600] in the second BNL analysis was mainly a result of the
analyzed $t'$~range.

A remaining puzzle is that although the \PpiOne[1600] decays into
$\Prho\pi$, it does not seem to be observed in
photoproduction.\cite{Nozar:2008aa,Eugenio:2013xua,Grabmuller:2012oja}
In the future, much more precise photoproduction data from the GlueX
and MesonX experiments at JLab will help to clarify the situation.

\subsection{$\eta\pi$ and $\eta'\pi$ Final States}

Other interesting final states to search for spin-exotic mesons are
$\eta\pi$ and $\eta'\pi$.  In contrast to the analysis of three-body
final states, where PWA models usually employ the isobar model thereby
introducing a large model dependence, the PWA of two-body final states
does not require such strong model assumptions.  In the \etaOrPrPi
system, partial waves that correspond to non-zero, odd orbital-angular
momentum between the~\etaOrPr and the~$\pi$ correspond to spin-exotic
quantum numbers.  The lowest such wave is the $P$-wave with
$\JPC = 1^{-+}$.  Previous experiments have observed the \PpiOne[1400]
in the $\eta\pi$ $P$-wave and the \PpiOne[1600] in the $\eta'\pi$
$P$-wave.\cite{Tanabashi:2018zz}

Recently, members of the Joint Physics Analysis Center (JPAC) have
performed a coupled-channel fit of the $\eta\pi$ and $\eta'\pi$ $P$-
and $D$-wave amplitudes extracted from COMPASS
data\cite{Adolph:2014rpp} using a unitary model based on $S$-matrix
principles.\cite{Rodas:2018owy}  They find two resonance poles, the
\PaTwo and the \PaTwo[1700], in the $D$-wave amplitudes and a single
pole in the $P$-wave amplitudes.  The pole parameters of
$m_0 = \SIerrs{1564}{24}{86}{\MeVcc}$ and
$\Gamma_0 = \SIerrs{492}{54}{102}{\MeVcc}$ are consistent with the
\PpiOne[1600].


\begin{wrapfigure}[21]{r}{0pt}
  \centering
  \includegraphics[width=0.34\textwidth]{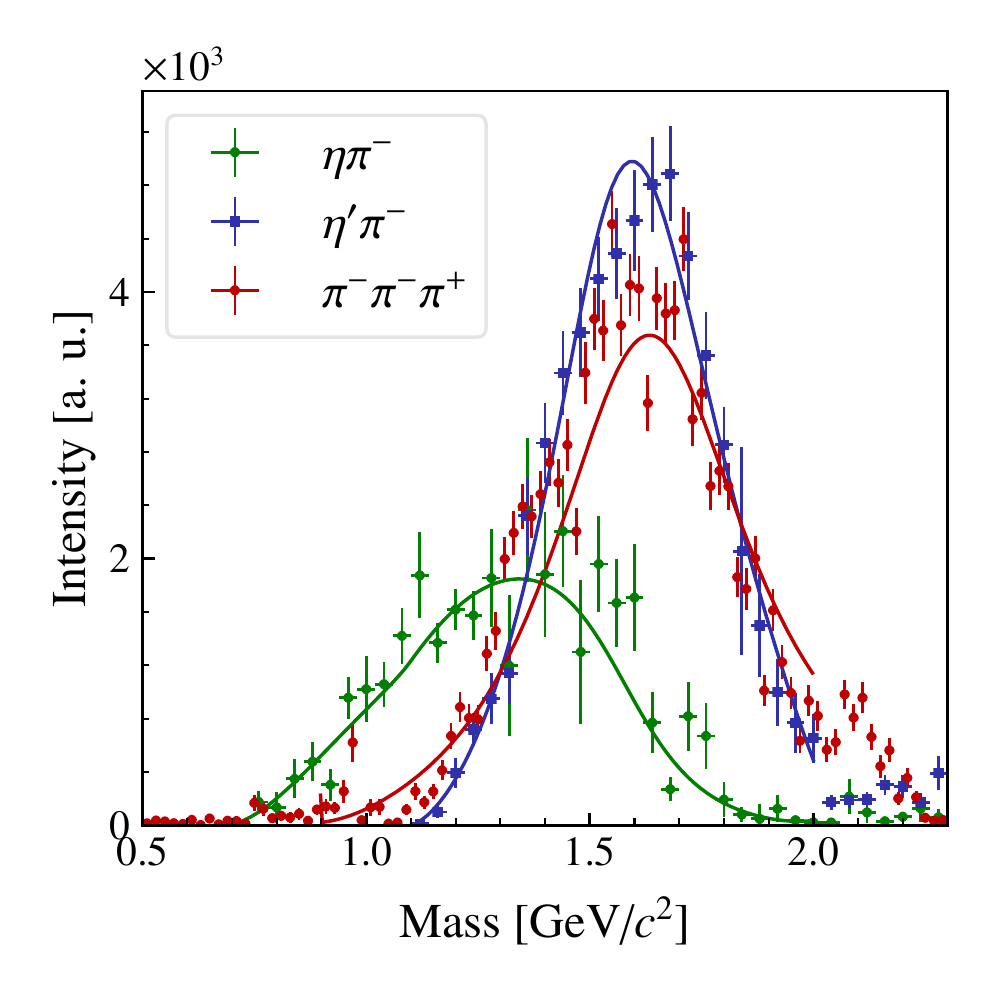}%
  \caption{Comparison of COMPASS data: Intensity distributions of the
    $\eta\pi$ (green points) and $\eta'\pi$ (blue points)
    $P$-wave\cite{Adolph:2014rpp} for \newline
    \SIvalRange{0.1}{t'}{1.0}{\GeVcsq} and of the $1^{-+}$ $\Prho\pi$
    $P$-wave in the \threePi data for
    \SIvalRange{0.449}{t'}{0.724}{\GeVcsq} (red points).  The curves
    represent the fit results from
    \refsCite{Rodas:2018owy,Akhunzyanov:2018lqa}.}
  \label{fig:1mp_compass}
\end{wrapfigure}

This result is in so far remarkable as only a single pole is required
to describe both the $\eta\pi$ and the $\eta'\pi$ $P$-wave amplitudes
despite their rather different intensity distributions (see
\cref{fig:1mp_compass}).  As mentioned above, this is in contrast to
most previous analyses, where the broad peak at \SI{1.4}{\GeVcc} in
the $\eta\pi$ $P$-wave intensity was described by the \PpiOne[1400]
whereas the narrower peak at \SI{1.6}{\GeVcc} in the $\eta'\pi$
$P$-wave intensity, which is nearly identical to the peak observed in
the high~$t'$ region of the \threePi data, was described by the
\PpiOne[1600].  As the COMPASS data are similar to those of previous
experiments, the JPAC analysis raises doubts about the existence of
the \PpiOne[1400] as a separate resonance.  This would resolve the
longstanding puzzle of two spin-exotic states lying unexpectedly close
to each other.  If interpreted as hybrid states, this would also
remove the discrepancy with lattice QCD and most model calculations,
which predict the lightest hybrid state to have a mass substantially
higher than that of the \PpiOne[1400] (see \eg\
\refCite{Meyer:2015eta}).

\subsection{GlueX Data}

The GlueX experiment at JLab\cite{Ghoul:2015ifw} uses a linearly
polarized photon beam to study photoproduction of light mesons on a
proton target.  The main goal is a precision measurement of the
light-meson spectrum in the mass range $\lesssim \SI{3}{\GeVcc}$ and
in particular the search for hybrid mesons.  The maximum photon
polarization is reached at a photon energy of about \SI{9}{\GeV}.  In
this energy range, various exchange processes contribute and states
with a wide variety of $\IGJPC$ quantum numbers\footnote{Here, $I$~is
  the isospin and $G$~the $G$-parity.} are accessible.  The photon
polarization helps to disentangle these production processes.  GlueX
is hence complementary to high-energy pion-scattering experiments such
as COMPASS, where only states with $\IG = 1^{-}$ are produced
directly.

\begin{figure}[tbp]
  \vspace*{-4ex}%
  \centering
  \subfloat[][]{%
    \includegraphics[width=0.5\textwidth]{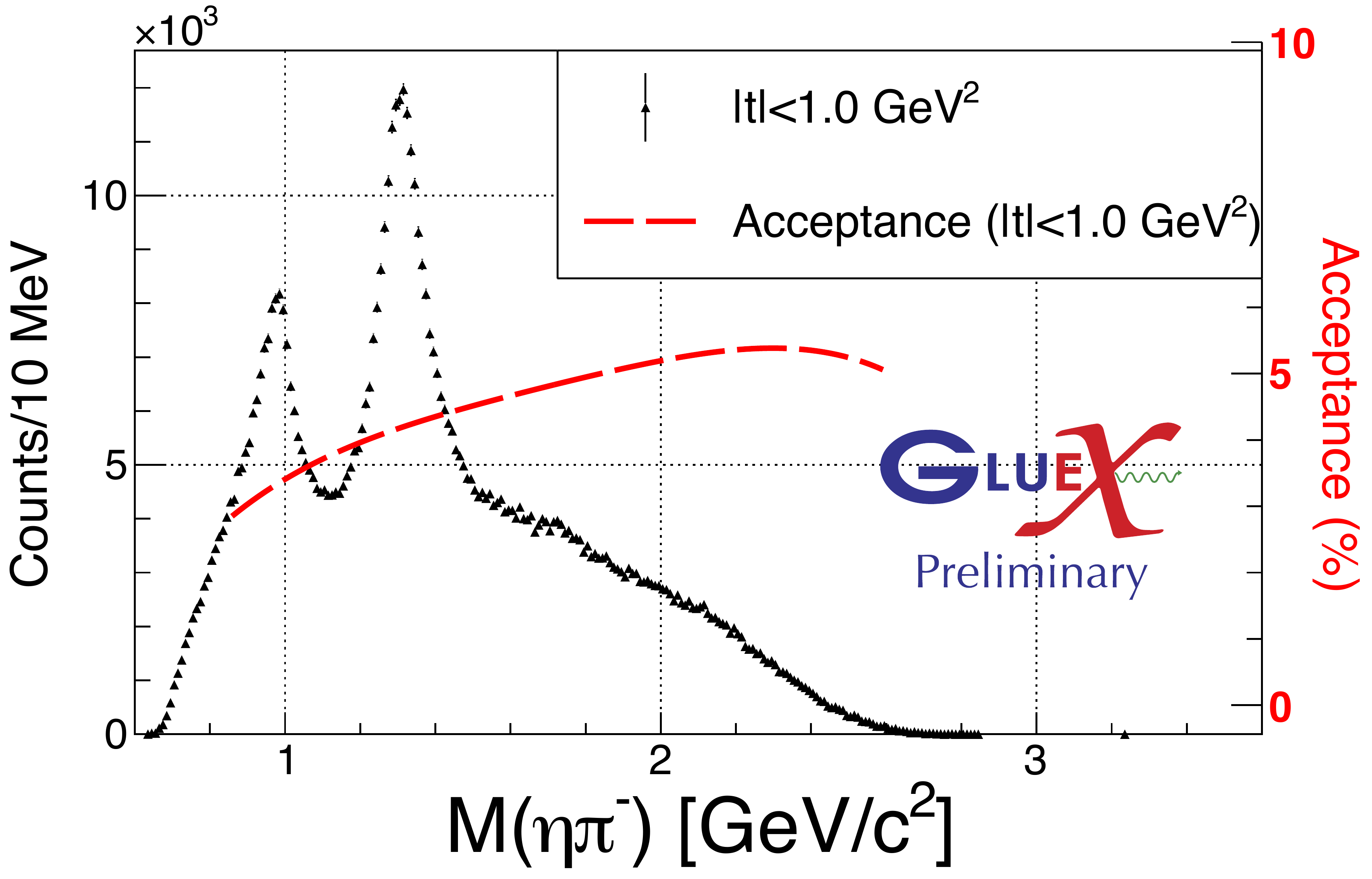}%
    \label{fig:etapi_mass_gluex}%
  }%
  \subfloat[][]{%
    \includegraphics[width=0.5\textwidth]{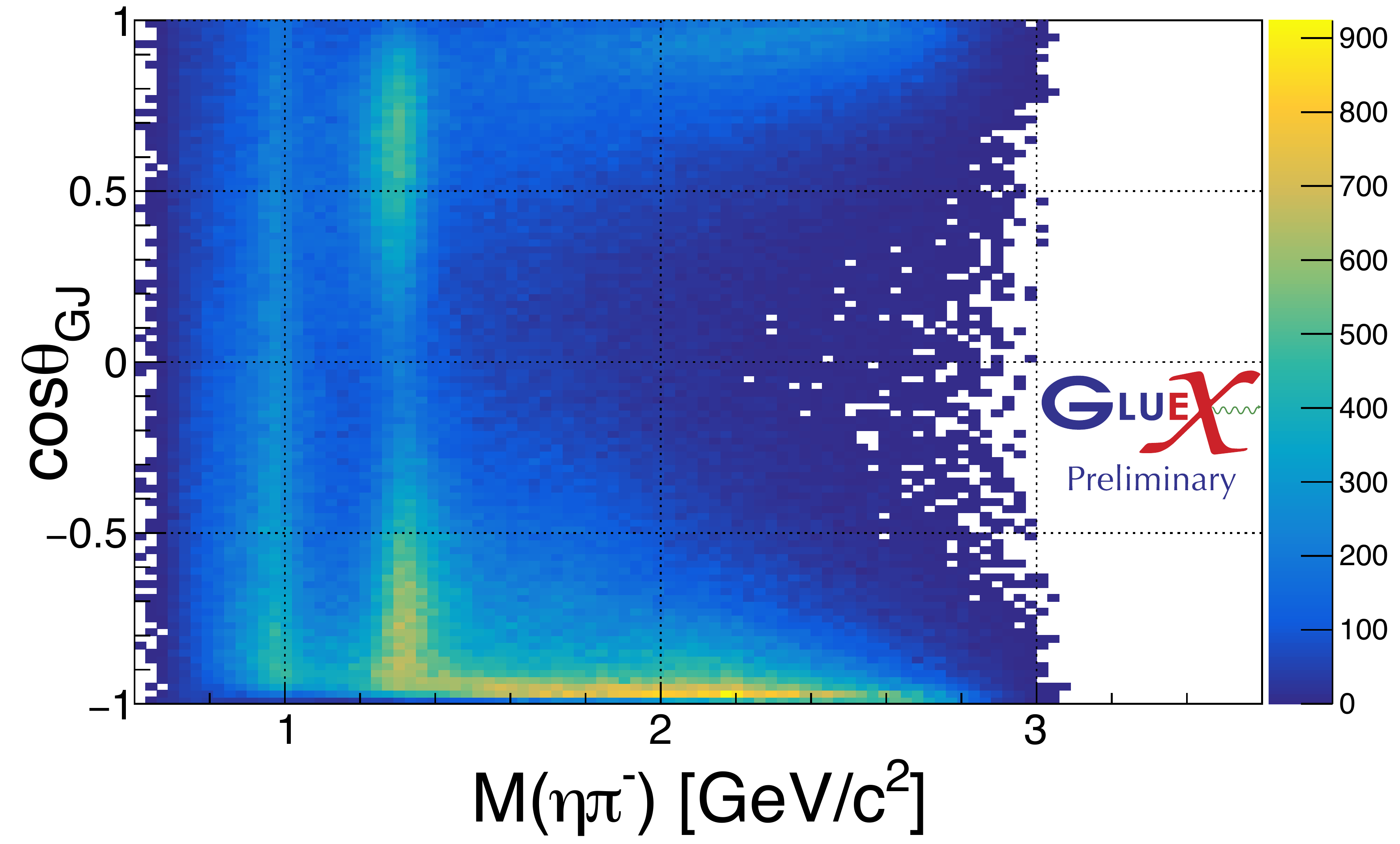}%
    \label{fig:etapi_mass_angle_vs_mass}%
  }%
  \caption{\subfloatLabel{fig:etapi_mass_gluex}~Invariant mass
    distribution of the $\eta\pi^-$ system produced in the reaction
    $\gamma p \to \eta\pi^- \Delta^{++}$ as measured by the GlueX
    experiment.\cite{Gleason:hadron2019}
    \subfloatLabel{fig:etapi_mass_angle_vs_mass}~Distribution of the
    cosine of the polar angle of the~$\eta$ \wrt the beam photon in
    the $\eta\pi$ rest frame vs. the $\eta\pi$ mass.}
  \label{fig:etapi_gluex}
\end{figure}

GlueX has finished its first phase of data taking and has acquired
world leading data samples for many final states.  First analyses
showcase the potential of these
data.\cite{Mack:hadron2019,Gleason:hadron2019}  As an example,
\cref{fig:etapi_gluex} shows kinematic distributions for the reaction
$\gamma p \to \eta\pi^- \Delta^{++}$ with $\eta \to \gamma \gamma$.
The $\eta\pi$ invariant mass distribution in
\cref{fig:etapi_mass_gluex} exhibits clear peaks of \PaZero and
\PaTwo.  This is also consistent with the angular distribution shown
in \cref{fig:etapi_mass_angle_vs_mass}.  The band in the \PaZero
region is approximately independent of the angle indicating a spin-0
resonance, whereas the \PaTwo region shows the expected 2-bump
behaviour.  In total, about \num{e6} $\eta\pi$ events are expected in
this channel, which is about 10~times the size of the COMPASS data
sample.  The second data-taking campaign of GlueX that started fall
2019 is expected to enlarge the data samples by another factor of
\numrange{4}{5}.

\section{Kaon Spectroscopy}

In order to better understand the light-meson spectrum it is important
to complete the SU(3)$_\text{flavor}$ nonets.  However, the kaon
spectrum is not well known.  Currently, the PDG lists only 25~kaon
states, 12~of which need confirmation.\cite{Tanabashi:2018zz}

\begin{figure}[tbp]
  \vspace*{-4ex}%
  \centering
  \subfloat[][]{%
    \includegraphics[width=0.333\textwidth]{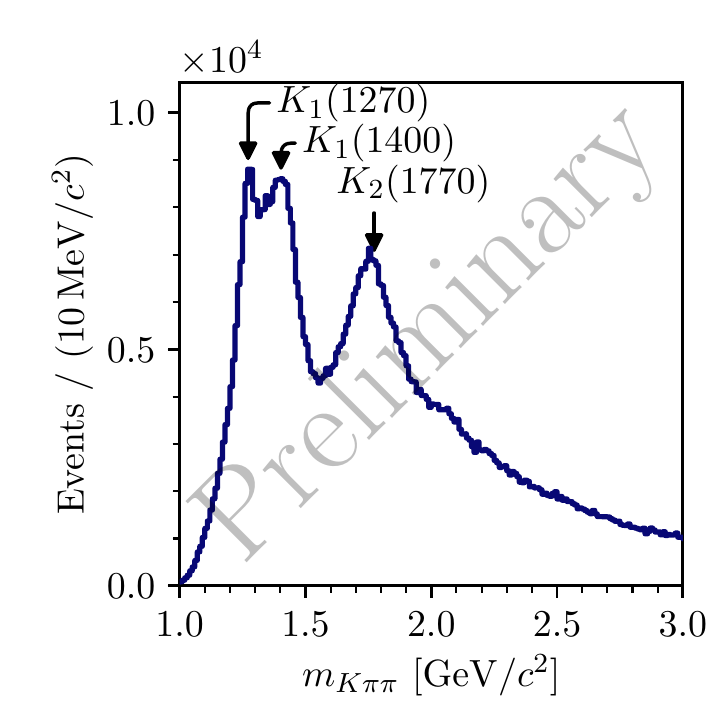}%
    \label{fig:kpipi_mass_compass}%
  }%
  \subfloat[][]{%
    \includegraphics[width=0.333\textwidth]{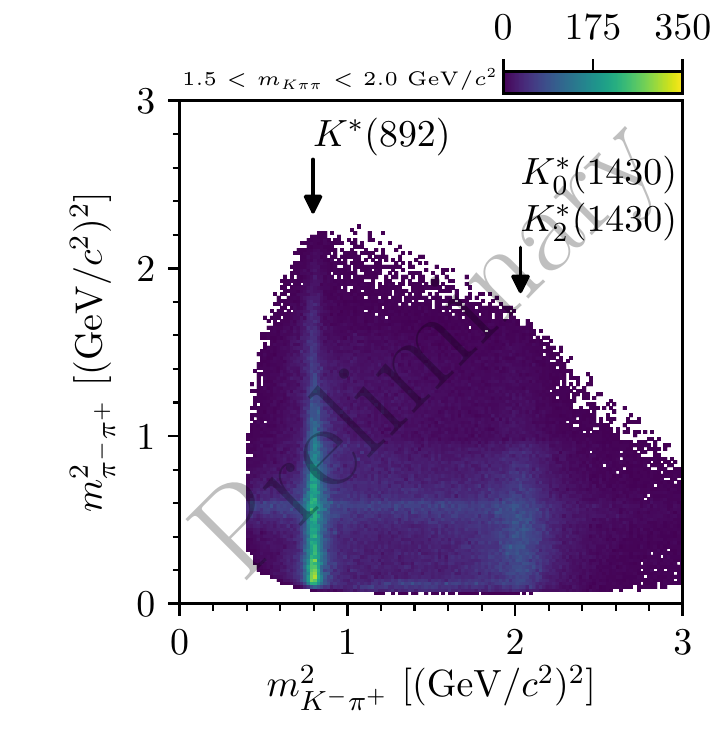}%
    \label{fig:kpipi_dalitz_compass}%
  }%
  \subfloat[][]{%
    \includegraphics[width=0.333\textwidth]{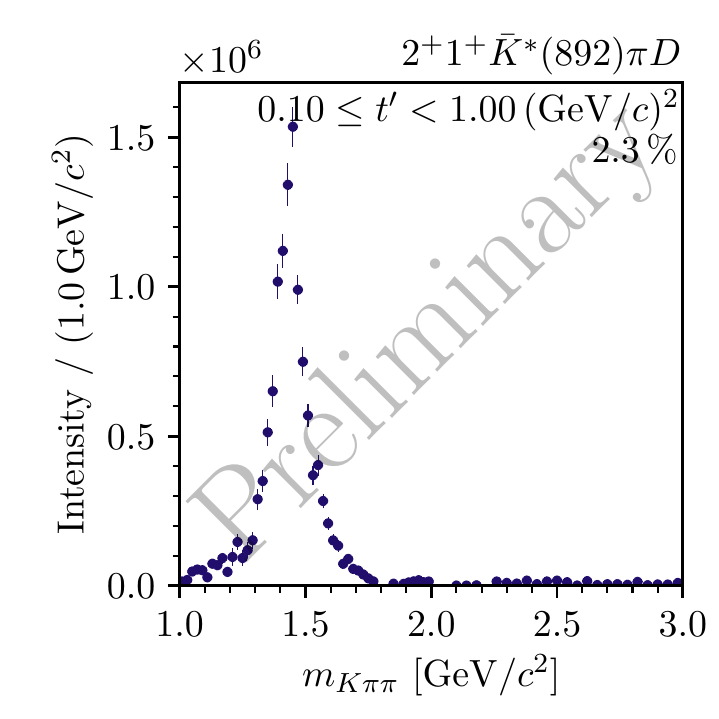}%
    \label{fig:kpipi_2+_int_compass}%
  }%
  \caption{\subfloatLabel{fig:kpipi_mass_compass}~\Kpipi invariant
    mass distribution with potential resonance signals indicated.
    \subfloatLabel{fig:kpipi_dalitz_compass}~Dalitz plot around
    $\mKpipi = \SI{1.75}{\GeVcc}$.
    \subfloatLabel{fig:kpipi_2+_int_compass}~Intensity distribution of
    the $\JP = 2^+$ wave decaying into \PKStar and $\pi^-$ in a
    $D$-wave.}
  \label{fig:kpipi_compass}
\end{figure}

The COMPASS experiment has measured kaon diffraction on a proton
target using the \SI{2.4}{\percent}~$K^-$ component in the
\SI{190}{\GeVc} hadron beam.  In particular, COMPASS has acquired the
so far largest data sample of \num{720000}~events of the reaction
$K^- p \to \Kpipi p$.\cite{Wallner:hadron2019}  The invariant mass
distribution of the \Kpipi system shown in
\cref{fig:kpipi_mass_compass} exhibits possible signals of known kaon
resonances.  The Dalitz plot in \cref{fig:kpipi_dalitz_compass} around
$\mKpipi = \SI{1.75}{\GeVcc}$ shows rich structures with signals of
\Prho and \PfZero[980] in the \twoPi subsystem and of \PKStar,
\PKZeroStar, and/or \PKTwoStar in the \Kpi subsystem.

A first PWA of these data was performed using a wave set determined by
employing regularization techniques.\cite{Kaspar:hadron2019}  To
exemplify the potential of the COMPASS data,
\cref{fig:kpipi_2+_int_compass} shows the intensity of the $\JP = 2^+$
wave decaying into \PKStar and $\pi^-$ in a $D$-wave, which exhibits a
nearly background-free signal of the \PKTwoStar.  More details can be
found in \refCite{Wallner:hadron2019}.

\section{Summary}

Light-meson spectroscopy has entered the era of high-precision data.
These data reveal new details of the light-meson spectrum and help to
settle some of the controversies of the past.  The currently running
GlueX and VES experiments and the soon to start MesonX experiment will
provide even more precise data on various final states.  In the more
distant future, the PANDA experiment will also contribute to this
field.

However, the PWA results from these large data samples become
increasingly dominated by systematic uncertainties.  A large
contribution to these uncertainties comes from the model assumptions
employed in the PWA of multi-body final states.  Reducing these
systematic uncertainties will require improved PWA models that respect
fundamental physical principles such as analyticity and unitarity (see
\eg\ \refCite{Mikhasenko:2017rkh}).  For scattering experiments, in
addition the production processes needs to be better understood in
order to improve the treatment of non-resonant processes, which are
another source of large systematic uncertainties.


\small
\bibliographystyle{ws-procs9x6} 
\bibliography{bgrube}

\end{document}